\documentclass[a4paper]{jpconf}
\usepackage{graphicx}
\usepackage{amsfonts}

\usepackage{amsmath}
\usepackage{amsthm}
\usepackage{amssymb}

\newcommand\be{\begin{equation}}
\newcommand\ee{\end{equation}}
\newcommand\bea{\begin{eqnarray}}
\newcommand\eea{\end{eqnarray}}

\newcommand\minus\backslash

\theoremstyle{definition}

\theoremstyle{remark}

\begin{document}
\title{Classical and quantum higher order superintegrable systems from coalgebra symmetry}

\author{D. Riglioni}
 
\address{Centre de Recherches Mat\'ematiques, Universit\'e de Montr\'eal}

\ead{ riglioni@crm.umontreal.ca}

\begin{abstract} The N-dimensional generalization of Bertrand spaces as families of Maximally superintegrable systems on spaces with nonconstant curvature is analyzed. Considering the classification of two dimensional radial systems admitting 3 constants of the motion at most quadratic in the momenta, we will be able to generate a new class of spherically symmetric M.S. systems by using a technique based on coalgebra.
The 3-dimensional realization of these systems provides the entire classification of classical spherically symmetric M.S. systems admitting periodic trajectories. We show that in dimension $N>2$ these systems (classical and quantum) admit, in general, higher order constants of motion and turn out to be exactly solvable. Furthermore it is possible to obtain non radial M.S. systems by introducing projection of the original radial system to a suitable lower dimensional space.    
\end{abstract}


\section{Introduction}
A Maximally Superintegrable (MS) system in classical mechanics is an integrable N-dimensional Hamiltonian system which 
is endowed with the maximum possible number of $2N-1$ functionally independent integrals of motion. 
In the last decades there has been a growing attention towards this class of systems due to their physically relevant properties both in classical and quantum mechanics :  they are conjectured to be exactly solvable \cite{tempturbwint} , they are often multiseparable  (the system can be separated in more than one coordinate system),   the (M.S.) classical systems exhibit periodic trajectories for all of their bounded motion and the quantum mechanical ones have degenerate spectrum for bound states. Because of these remarkable properties we find (M.S.) systems as successful models in many areas of physics such as in condensed matter physics as well as atomic, molecular and nuclear physics see e.g. \cite{quesne} \cite{feher}   \cite{gritsev} and reference therein.
Superintegrable systems are very rare, in the sense that, under certain assumptions  it is possible to classify completely all the possible systems in the (M.S.) family : The most celebrated example dates back to the 19th century and is a consequence of the Bertrand's theorem \cite{Be73}. Given a Hamiltonian system defined on a flat space, then the only central potentials admitting the (M.S.) property are 

$$
H= p_r^2 + \frac{p_\theta^2}{r^2} + V(r), \quad V(r) = \{\frac{\mu}{r}, \omega^2 r^2 \},
$$

\noindent the harmonic oscillator or Kepler-Coulomb potential. These two systems are also characterized by having quadratic constants of the motion  $I^{(2)}$
$$ \{H,I \}=0  : I(r,\theta,p_r,p_\theta)^{(N)} = \sum_{n=0}^N \sum_{i=0}^n a(r,\theta)_{n,i} p_r^{n-i} p_\theta^i, \quad N=2  .$$

Over the last century many other classifications have been made, above all considering superintegrability characterized by constants of the motion $I^{(N)}$ of a fixed order $N$ defined
on both Euclidean and non Euclidean spaces. The literature about it is simply too vast to cite all the important contributions, however a quite exhaustive list can be found in the review paper \cite{marquette} where superintegrable systems with constants of the motion up to the third order are considered.
The main goal of the paper is analyzing the classification of all the M.S. systems on general N-dimensional Non-Euclidean spaces characterized by radial symmetry and admitting bound states, which, in the 3-dimensional case coincide with the so called Perlick systems \cite{Pe92}.
This class of systems has already been introduced in a series of papers by Ballesteros et al. and the author  \cite{Betal1} \cite{Betal2} \cite{Betal3} \cite{Rig} in which the Maximal Superintegrability of all the radial systems defined on Perlick's space-times is proven in their classical version and for few of them also in the quantum one. 
In this paper we will present a strategy to obtain systematically all the constants of the motion $I^{(N)}$ (which in general are of higher order $N$) for these systems  in spaces of arbitrary dimension both in classical and quantum mechanics.  
In doing this we will emphasize the special role played by the systems admitting quadratic constants of the motion $H, I^{(2)}$: The classification of quadratic (M.S.) systems with radial symmetry on  two dimensional non-Euclidean spaces admitting periodic trajectories can be expressed in terms of two families which can be regarded as deformations of the Kepler and Harmonic oscillator systems \cite{Betal4} \cite{Betal5}.

$$
H_I = \frac{(1 + k r^2)^2}{2} (p_r^2 + \frac{p_\theta^2}{r^2}) + A \frac{1 - k r^2}{r}
$$ 
$$
H_{II} = \frac{(1 - \lambda^2 r^4)^2}{2 (1 + \lambda^2 r^4 - 2\delta r^2)} (p_r^2 + \frac{p_\theta^2}{r^2}) + \frac{B r^2}{ 1 + \lambda^2 r^4 -2 \delta r^2}.
$$   
  
\noindent We  will show how the canonical transformation  

$$
r = r'^{\beta} ;P_r = \frac{r'^{1 - \beta}}{\beta} p_{r'} ; 
$$
$$
\theta = \beta \theta' ; p_\theta = \frac{p_{\theta'}}{\beta}
$$
introduces a new angular parameter $\beta$ which, after a proper embedding procedure into a 3 dimensional space $p_\theta^2 \rightarrow p_\theta^2 + \frac{p_\phi^2}{\sin^2 \theta}$, turns the 2-D quadratic systems into  
 
$$
H_I = \frac{r'^2 (r'^{- \beta} + k r^\beta)^2}{2 \beta^2 } (p_{r'}^2 + \frac{p_{\theta'}^2}{r'^2} + \frac{p_{\phi'}^2}{r'^2 \sin^2 \theta'}) + A (r'^{-\beta} -k r^{\beta}),
$$ 
$$
H_{II} = \frac{r'^2 (r'^{-2 \beta} - \lambda^2 r'^{2 \beta})^2}{2 (r'^{-2 \beta} + \lambda^2 r'^{2 \beta} - 2\delta )} (p_{r'}^2 + \frac{p_{\theta'}^2}{r'^2} + \frac{p_{\phi'}^2}{r'^2 \sin^2 \theta'})+ \frac{B }{ r'^{-2 \beta} + \lambda^2 r^{2 \beta} -2 \delta }.
$$  

\noindent This is the entire Perlick family, which in contrast to the two dimensional case, is characterized by having in general higher order constants of the motion $I^{(N)}$.
As we will show in section (\ref{quattro}) this fact can be understood in terms of the $sl2$ coalgebra symmetry \cite{Betal6} which is common to all the Perlick Hamitlonians regardless the dimension of the space in which they are embedded. This is in contrast with the representation of its algebra generators, which generally depends on the dimension of the space we are considering and that, in the particular case of a 2-dimensional space, admits always a quadratic representation for all the first integrals.  
\newpage 
\noindent Moreover the identification of the coalgebra structure for these systems opens a way to obtain a wider class of higher order superintegrable systems by considering just different representations 
of the same algebra. As a consequence of this we will show explicitly how it is possible, by choosing different representations of the coalgebra elements, to transform the family of Perlick systems into the family of the well known TTW systems and its extensions \cite{TTW} \cite{PW} \cite{miller}. These exhibit higher order constants of the motion on the same footing of the Perlick systems.
To conclude the paper we will discuss the quantum version of all these systems stressing the role of the SUSY QM (Super Symmetric Quantum Mechanics) that can be fruitfully used to get the solution and the superintegrable quantization of the MS Hamiltonians and its first integrals. 
The paper is organized as follows:
In Section 2 we will give a survey of radial superintegrable systems on two dimensional non-Euclidean spaces. In Section 3 we will introduce the 3-dimensional Perlick systems stressing their connections with the classifications of the quadratic two dimensional superintegrable systems introduced in the Section 2. In Section 4 we will analyze the integrability properties of the Perlick systems in terms of the coalgebra $sl2$. In Section 5 we will extend the coalgebra approach to obtain non-radial higher order M.S. systems. Finally in Section 6 and 7 we will provide the quantum version of all the systems previously analyzed in sections (1-5).

\section{Two dimensional Maximal Superintegrable radial systems}
It is very well known that for any classical N-dimensional radial Hamiltonian system all the trajectories describe a planar motion. Their quantum analogues have a spectrum which depends on just two quantum numbers. This fact can be explained in terms of superintegrablility claiming that every N-dimensional radial system has a "coalgebra-symmetry" \cite{Betal6} which makes the system itself quasi-maximally superintegrable, namely it has at least $2N-2$ integrals of motion whose main effect is that of reducing its dynamics to that of a 2 dimensional system. In the light of these considerations let us start our analysis giving a brief review of the classification of two dimensional M.S. systems. In particular we will show how the entire family of radial M.S. systems admitting periodic trajectories can be obtained from a special class of M.S. systems which is characterized by the fact of having at most quadratic integrals of the motion. Let us introduce a general two-dimensional Riemannian space whose metric is given by:

\begin{equation}
ds^2 = g_{i,j} dx^i dx^j, \quad i,j =1,2 
\end{equation}

The Hamiltonian describing a particle moving on such a space and subjected to a potential $V(x^1,x^2)$  has the following form for a classical system:

\begin{equation}
\label{hamiltonclas}
H = \frac{1}{2} g^{i,j} p_i p_j + \mu V(x_1,x_2),
\end{equation}

while for a quantum mechanical system we have:

\begin{equation}
\label{hamiltonquant}
\hat{H} = -\frac{1}{2 \sqrt{g}} \partial_i (\sqrt{g} g^{i,j} \partial_j ) +\mu V(x_1,x_2).
\end{equation}

The form of the Hamiltonians introduced above induces a preliminary discussion on what are the possible Riemannian spaces admitting a M.S. Hamiltonian system. If we exclude from this discussion the trivial ones like the Euclidean space and the constant curvature space then the analysis reduces to the classification of all those Riemannian spaces with a non constant scalar curvature whose associated Hamiltonian admits three functionally independent constants of the motion. A significant result in this sense was obtained by Koenigs \cite{koenigs} in 1889. He classified all those Riemannian spaces whose Hamiltonian admits three constants of the motion $I$ at most quadratic in the momentum. 
\newpage 
\noindent More explicitly such a condition in classical and quantum mechanics is respectively equivalent to:

\begin{equation}
I = a(x_i,x_j)^{i,j}p_i p_j  +  f(x_1,x_2): \quad \{H,I\} = 0 
\end{equation}
\begin{equation}
\hat{I} = a(x_i,x_j)^{i,j} \partial_i \partial_j + b(x_1,x_2)^i \partial_i + f(x_1,x_2) : \quad [\hat{H},\hat{I} ] = 0.
\end{equation}

\noindent Where the curly bracket $\{,\}$ stands for the Poisson brackets and the square one $[,]$ for the Lie commutator. The list of spaces which satisfy at these constraints are called Darboux spaces. This list contains 4 spaces which we report here in diagonal form as introduced in \cite{KKMW1} \cite{KKMW2} :

\begin{align}
(I) & &  2x (dx^2 + dy^2) \\
(II) & & \frac{x^2+1}{x^2} (dx^2 + dy^2) \\
(III) & & \frac{a e^{2x} + b}{e^{4x}} (dx^2 + dy^2) \\
(IV) & & - \frac{2 a \cosh(2x) + b}{4 \sinh^2 (2x)}  (dx^2 + dy^2). 
\end{align}

\noindent These metric spaces can be recast in a radial conformally flat reference frame by defining  the  transformation $(x,y) \rightarrow (r= \frac{e^{\epsilon x}}{\sqrt{\lambda}} , y = \theta)$, $\epsilon = \pm 1$. The Riemannian metric trasforms into $f(x) (dx^2 + dy^2) \rightarrow \frac{f(x(r))}{r^2} (dr^2 + r^2 d \theta^2)$  
 
\begin{align}
(I) & &  \frac{2 \epsilon \ln (\sqrt{\lambda_1} r)}{r^2} (dr^2 + r^2 d\theta^2) \\
(II) & & \left( \frac{1}{r^2} + \frac{1}{(r \ln (\sqrt{\lambda_2} r))^2} \right) (dr^2 + r^2 d\theta^2) \\
(III) & & \frac{a \lambda_3^\epsilon + b r^{-2 \epsilon}}{\lambda_3^{2 \epsilon} r^{2 \epsilon +2}} (dr^2 + r^2 d\theta^2) \\
(IV) & & - \frac{a \lambda_4^{3 \epsilon} r^{2 \epsilon} + a \lambda_4^\epsilon r^{- 2\epsilon} + b \lambda_4^{2 \epsilon}}{r^2 (\lambda_4^{2 \epsilon} r^{2 \epsilon} - r^{-2 \epsilon})^2 }  (dr^2 + r^2 d\theta^2) .
\end{align}

\noindent Let us note that the Darboux III and Darboux IV can be described as particular cases of a more general Riemannian space : 

\begin{equation}
\label{perlick}
ds^2 = \frac{2c (r^{-2 \epsilon} + \lambda^2 r^{2 \epsilon} - 2 \delta)}{r^2 (r^{- 2\epsilon} - \lambda^2 r^{2 \epsilon})^2 } (dr^2 +r^2 d\theta^2). 
\end{equation}

\noindent  It is straightforward to see that the Darboux IV metric coincides with  (\ref{perlick}) if we set the parameters equal to  $c = \frac{-a \lambda_4^\epsilon}{2}, \delta = \frac{-b \lambda_4^\epsilon}{2a}, \lambda = \lambda^\epsilon_4 $. The previous change of parameters is not defined for $\lambda = 0$. This limit describes the Darboux III space with parameters $c = \frac{a}{2 \lambda_3^\epsilon}, \delta = \frac{-b}{2a \lambda_3^\epsilon},\lambda=0, \epsilon \rightarrow - \epsilon$. For each family of M.S. spaces the second step consists in seeing if the Hamiltonians \ref{hamiltonclas} and \ref{hamiltonquant} defined on the Darboux spaces admit a potential $V$ which keeps the quadratic superintegrability of the system. An answer to this question was provided more recently by \cite{KKMW1} \cite{KKMW2} who classify the M.S. potentials which we can consider without breaking the quadratic superintegrability of the free motion on Darboux spaces. What turned out from that classification was that all of the M.S. Darboux systems with potential can be obtained through the Coupling Constant Metamorphosis (CCM) applied to M.S. systems on spaces of constant curvature.To be selfcontained let us  recall briefly what the CCM is.  For our purpose the coupling constant metamorphosis can be briefly summarized as follows.
Let us consider $T$ as the kinetical energy term given by $T = \sum_{i,j} g^{i,j} p_i p_j$ and $U$, $V$ are potentials terms independent of the arbitrary parameter $\mu$.
$$
H = T + V -\mu U.
$$
 Let $S=S(\mu)$ be an integral of the motion. We define the action of the coupling constant metamorphosis on $H$ and $S$ as :

\begin{equation}
\tilde{H} = \frac{1}{U} (T + V - E), \tilde{S} \equiv S(\tilde{H}),
\end{equation}

then $\tilde{S}$ is an integral of the motion for $\tilde{H}$.  Let us apply this machinery to the generalization of the Kepler problem on a space of constant curvature. The Kepler problem on a two dimensional space of constant curvature in cartesian and polar coordinates is given by:

\begin{equation}
H = \frac{(1+k(x^2+y^2))^2}{2} (p_x^2 + p_y^2) - \mu \frac{1-k(x^2+y^2)}{\sqrt{x^2+y^2}} +4 \mu \delta 
\end{equation}  

\begin{equation}
\label{keplers}
H= \frac{(1+k r^2)^2}{2} (p_r^2 + \frac{p_\theta^2}{r^2}) - \mu \frac{1-kr^2}{r} +4 \mu \delta 
\end{equation}

let us consider the Levi Civita transformation :

\begin{equation}
\begin{cases}
x = \frac{\tilde{x}^2- \tilde{y}^2}{2} ; r= \frac{\tilde{r}^2}{2}\\
y=\tilde{x}\tilde{y}; \theta = 2 \tilde{\theta} 
\end{cases}
\end{equation}

The Hamiltonian in the new cartesian coordinates is 

\begin{equation}
H = \frac{((\tilde{x}^2 + \tilde{y}^2)^{-1} - \lambda^2(\tilde{x}^2 + \tilde{y}^2))^2 (\tilde{x}^2 + \tilde{y}^2)}{2} (p_{\tilde{x}}^2 + p_{\tilde{y}}^2 ) - \tilde{\mu} ((\tilde{x}^2 + \tilde{y}^2)^{-2} + \lambda^2 (\tilde{x}^2 + \tilde{y}^2)^{2} - 2 \delta) 
\end{equation}

or in polar coordinates

\begin{equation}
\label{darboux}
H = \frac{(r^{-2} - \lambda^2 r^2)^2 r^2}{2} (p_r^2 + \frac{p_\theta^2}{r^2}) - \tilde{\mu} (r^{-2} + \lambda^2 r^2 -2 \delta)
\end{equation}

where $\lambda^2 = -\frac{k}{4}$ , $\mu = \frac{\tilde{\mu}}{2}$ . \newline
Let us apply the coupling constant metamorphosis to this system. The potential $U$ turns out to be
 $$
U = (r^{-2} + \lambda^2 r^2 -2 \delta)
$$

and the Hamiltonian $\tilde{H}$ is

\begin{equation}
\label{darbouxiiquad}
\tilde{H} = \frac{(r^{-2} - \lambda^2 r^2)^2 r^2}{2(r^{-2} + \lambda^2 r^2 -2 \delta)} (p_r^2 + \frac{p_\theta^2}{r^2}) + \frac{E}{r^{-2} + \lambda^2 r^2 -2 \delta}.
\end{equation}

It is straightforward to see that the kinetical part of the CCM Hamiltonian $\tilde{H}$ turns out to  describe a particle moving on a Darboux type space \ref{perlick} when  $\epsilon = 1$, moreover the CCM provides directly also the family of possible M.S. radial potentials admitting bound states on Darboux spaces in full agreement with the classification given in \cite{KKMW1} \cite{KKMW2} .

\section{Beyond the quadratic M.S. and Bertrand systems}

In the previous section we have introduced a class of 2-dimensional M.S. systems admitting three constants of the motion at most quadratic in the momentum in Euclidean and Non Euclidean spaces.
As previously said the M.S. entails, for a classical Hamiltonian system, that all of its bounded trajectories are periodic, and this is indeed the case for the systems (\ref{keplers}, \ref{darbouxiiquad}) which describe elliptical trajectories.  Reversing the problem we can look for superintegrability among systems admitting stable periodic trajectories.
The idea of classifying Hamiltonian systems for which all bounded trajectories are closed dates back to  Bertrand theorem \cite{Be73} proving that the only radial potentials with this property are the harmonic oscillator and the Kepler potentials. That theorem has been recently generalized to non Euclidean radial spaces by Perlick in his remarkable paper \cite{Pe92}.  Analogously to the original Bertrand's theorem, Perlick found only two couples of (metric spaces, potentials ) admitting periodic trajectories for any bounded motion, which can be considered as the non-Euclidean deformation of the Kepler and Harmonic oscillator on a flat space.  Let us report these two families  in a conformally flat reference frame as earlier obtained in \cite{Betal5}. 

\begin{equation}
\label{coppia1}
ds_I^2 = \frac{1}{r^2(r^{-\beta} + k r^{\beta})^2} (dr^2 + r^2 d \theta^2 + r^2 \sin^2 \theta d \phi^2 ); \quad V_I = - \mu (r^{-\beta} -k r^{\beta}) , \quad \beta \in \mathbb{Q}
\end{equation}
  
\begin{equation}
\label{coppia2}
ds_{II}^2 = \frac{r^{-2 \gamma} + \lambda^2 r^{2 \gamma} -2 \delta}{r^2(r^{-2 \gamma} - \lambda^2 r^{2 \gamma})^2} (dr^2 + r^2 d \theta^2 + r^2 \sin^2 \theta d \phi^2 ); \quad V_{II} = \frac{\mu}{(r^{- 2 \gamma} + \lambda^2 r^{2 \gamma} - 2 \delta)}, \gamma \in \mathbb{Q} .
\end{equation}

The associated Hamiltonian systems turn out to be:

\begin{equation}
\label{perlickham1}
H_I = \frac{r^2(r^{-\beta} + k r^{\beta})^2}{2} \left( p_r^2 + \frac{L^2}{r^2} \right)  - \mu (r^{-\beta} -k r^{\beta}) 
\end{equation}

\begin{equation}
\label{perlickham2}
H_{II} = \frac{r^2(r^{-2\gamma} - \lambda^2 r^{2\gamma})^2}{2(r^{-2 \gamma} + \lambda^2 r^{2 \gamma} -2 \delta)} \left( p_r^2 + \frac{L^2}{r^2} \right) +  \frac{\mu}{(r^{- 2 \gamma} + \lambda^2 r^{2 \gamma} - 2 \delta)} 
\end{equation}
 
where $L^2 = p_\theta^2 +\frac{p_\phi^2}{\sin^2 \theta}, \quad (\beta, \gamma) = \frac{m}{n}; m,n \in \mathbb{N} $.
\newline As expected for Hamiltonian systems admitting periodic trajectories the systems (\ref{perlickham1}, \ref{perlickham2}) are M.S. with integrals of motion polynomial in the momentum of arbitrary order $r = m+n$ as proven in \cite{Betal1}, and define implicitly the most general classification of classical M.S. systems with radial symmetry admitting periodic trajectories (namely bound motion). This classification of M.S. systems overlaps with the previous classfication of quadratic M.S. systems on two dimensional spaces whenever we fix the plane of the motion.
Let us consider the Hamiltonian systems (\ref{perlickham1}, \ref{perlickham2}) in the case $p_\phi = 0$. This choice makes the Hamiltonians (\ref{perlickham1}, \ref{perlickham2}) comparable with the two dimensional Hamiltonians obtained in (\ref{keplers}, \ref{darbouxiiquad}), and in fact they coincide with the (\ref{perlickham1}, \ref{perlickham2}) if we consider the particular case $\beta=1$, $\gamma =1$.
\newline {\bf{Remark}}
\newline The choice of a particular plane of the motion is not restrictive in order to analyze the dynamics of a radial system, nevertheless the reduction of a 3-dimensional system to a two-dimensional one changes deeply the geometric properties of the space, and in fact the metric space (\ref{coppia1}) has a scalar curvature which turns out to be in general non constant :

\begin{equation}
R_I = 2(1-\beta^2) (r^{-\beta} +k r^{\beta})^2 + 24 \beta^2 k
\end{equation}     
 
while if we consider a fixed $\phi$ the reduced first fundamental form 

\begin{equation}
ds^2 =  \frac{1}{r^2(r^{-\beta} + k r^{\beta})^2} (dr^2 + r^2 d \theta^2 )
\end{equation}   

has associated a scalar curvature which is constant for each value of $\beta$ 

\begin{equation}
R_{Ir} =  8 \beta^2 k.
\end{equation}     

The main consequence of this fact is that the two dimensional reduction of (\ref{perlickham1})
can be recast in (\ref{keplers}) through a canonical change of variables 

\begin{equation}
\label{levicivitageneral}
\begin{cases}
r = r'^{\frac{1}{\alpha}}  ; p_r= \alpha \frac{r'}{r'^{\frac{1}{\alpha}}} p_{r'}\\
\theta =\frac{\theta'}{\alpha}; p_\theta = \alpha p_{\theta'} 
\end{cases}
\quad \alpha = \beta = \gamma
\end{equation} 

\begin{equation}
H = \beta^2 \frac{r'^2 (r'^{-1} + k r')^2}{2} \left( p_{r'}^2 + \frac{p_\theta^2}{r'^2} \right)  - \mu (r'^{-1} -k r') .
\end{equation}
The same transformation (\ref{levicivitageneral}) can be used to recast the (\ref{perlickham2}) into the (\ref{darbouxiiquad}) 

\begin{equation}
H = \gamma^2 \frac{r'^2(r'^{-2} - \lambda^2 r'^{2})^2}{2(r'^{-2} + \lambda^2 r'^{2} -2 \delta)} \left( p_{r'}^2 + \frac{p_{\theta'}^2}{r'^2} \right) +  \frac{\mu}{(r'^{- 2} + \lambda^2 r'^{2} - 2 \delta)} .
\end{equation}

This entails that the quadratic M.S. systems on space of constant curvature and its coupling constant metamorphosis partners describe the dynamic of the whole family of radial systems whose bound trajectories are periodic.   

\section{Higher order constants of the motion for Perlick's Hamiltonians}
\label{quattro}
The above considerations suggest a connection between the quadratic constants of the motion of (\ref{keplers}) and (\ref{darbouxiiquad}) and the constants of the motion of the Perlick's family which, generally are a polynomial of degree n in the momentum variables. Let us analyze this point computing all the constants of the motion for the system (\ref{keplers}).

\subsection{A "Laplace-Runge-Lenz vector" for the quadratic M.S. systems}

As previously stressed, the system (\ref{keplers}) plays a crucial role in the classification of M.S. systems with a radial symmetry: in the previous section we have seen how the dynamic of the system (\ref{perlickham1}) reduces to the dynamic of (\ref{keplers}) through a canonical change of variables, moreover the whole second family (\ref{perlickham2}) can be obtained through a coupling constant metamorphosis of (\ref{keplers}). The aim of this section is to analyze the constants of the motion associated to (\ref{keplers}) and to undersatand how the transformation (\ref{levicivitageneral}) can generate the higher order constants of the motion associated to (\ref{perlickham1} , \ref{perlickham2}).
The two dimensional Kepler system on a space of constant scalar curvature 

\begin{equation}
\label{sistbase}
H =  \frac{r'^2 (r'^{-1} + k r')^2}{2} \left( p_{r'}^2 + \frac{p_\theta^2}{r'^2} \right)  - \mu (r'^{-1} -k r') + 4 \mu \delta ,
\end{equation}

is M.S. since it has three constants of the motion: the Hamiltonian itself, the angular momentum coming from the radial symmetry $p_\theta$, and a "Laplace-Runge-Lenz" vector. Let us compute the "Laplace-Runge-Lenz" vector from the solution of the orbit equation.
\newline The orbit equation associated to the system (\ref{sistbase}) is 

\begin{equation}
d \theta = \frac{L dr}{r^2 \sqrt{\frac{2}{(1+kr^2)^2}\left( E + \mu \left( \frac{1-k r^2}{r} -4 \mu \delta \right) \right) - \frac{L^2}{r^2}}}, \quad L=p_\theta
\end{equation}  

 this can be recast in the form

\begin{equation}
d \theta = \frac{-L (- \frac{1+kr^2}{r^2}dr)}{\sqrt{2E + 2\mu \frac{1-k r^2}{r} -8 \mu \delta - L^2\left( \frac{(1-kr^2)^2}{r^2} + 4 k \right)}}
\end{equation}

the equation symplifies with the change of variables:

\begin{equation}
u = \frac{1- kr^2}{r}
\end{equation}

\begin{equation}
d \theta = \frac{- L du}{\sqrt{2E +2 \mu u - 8 \mu \delta -L^2 u^2 -4 k L^2 } } 
\end{equation}

which can be readily integrated to yield:

\begin{equation}
\label{orbiteq}
\cos(\theta - \theta_0) = \frac{L^2 \frac{1-kr^2}{r}-\mu}{\sqrt{2EL^2 - 8 \mu \delta L^2 -4 k L^4 + \mu^2}}
\end{equation}

$$
\sin( \theta - \theta_0 )= \frac{(1+kr^2) L p_r}{\sqrt{2EL^2 - 8 \mu \delta L^2 -4 k L^4 + \mu^2}}.
$$

The orbit periodicity induces the following complex "Laplace-Runge-Lenz" constant of the motion

\begin{equation}
\label{runge}
S = const \quad \!\!\! e^{i \theta(r,p_r,H,p_\theta)} e^{- i \theta} = \left( L^2 \frac{1-k r^2}{r} - \mu + i (1+kr^2) p_r L \right) e^{- i \theta}
\end{equation}

\noindent where $const = \sqrt{2EL^2 - 8 \mu \delta L^2 -4 k L^4 + \mu^2}$ is a constant of the motion since it is a function of $E,L$ . Both the Real and Imaginary part of (\ref{runge}) are constants of the motion quadratic in the momentum in radial variable and in cartesian ones, with $L = x p_y - y p_x$, $p_r = \frac{x p_x + y p_y}{\sqrt{x^2+y^2}}$ :

\begin{equation}
\label{cartesianrunge}
S = const \quad \!\!\! e^{i \theta(r,p_r,H,p_\theta)} e^{- i \theta} = \left( L^2 \frac{1-k r^2}{r} - \mu + i (1+kr^2) p_r L \right) \frac{x - i y}{r}.
\end{equation}

\subsection{embedding a two-dimensional radial system in a higher dimensional space }
\label{coproductsec}
The two dimensional M.S. system (\ref{keplers}) analyzed above can be embedded in a higher dimensional space without changing its integrability properties.
This operation can be performed if we look at the system (\ref{keplers}) and its constants of the motion (\ref{cartesianrunge} , $p_\theta$ ) as elements of a universal enveloping algebra of a Poisson-Lie algebra whose two dimensional realization can be regarded as just a particular choice.
Let us introduce the following $sl(2)$ Lie coalgebra 

\begin{equation}
\label{sl2}
\{ J_3 , J_+ \} = 2 J_+, \quad \{ J_3, J_- \} = -2 J_-, \quad \{ J_- , J_+ \}= 4 J_3
\end{equation} 

equipped with the trivial coproduct $\Delta$

\begin{equation}
\label{dcoproduct}
\Delta(1) = 1 \otimes 1 \quad \Delta(J_i) = J_i \otimes 1 + 1 \otimes J_i \quad i=+,-,3
\end{equation}

the action of the coproduct defines a homomorphism for the algebra

\begin{equation}
\{ \! \Delta( \! J_3 \!) ,   \Delta( \! J_+ \!) \! \} = 2 \! \Delta( \! J_+ \! ); \{ \! \Delta( \! J_3 \!)  ,  \Delta( \! J_- \! ) \! \} =  -2 \! \Delta( \! J_- \! ); \{ \! \Delta( \! J_- \! )  ,  \Delta( \! J_+ \! ) \! \}  =  4 \! \Delta( \!J_3 \!)
\end{equation}

Let us define moreover $\Delta()^n$ as the iteration of the coproduct $\Delta()^n = \underbrace{\Delta(\Delta(....))}_{\times n}$, then if we fix the maximum value of $n$ to $n \leq N$ we recover a Lie algebra which can be decomposed to a direct sum of algebra (\ref{sl2}) $\underbrace{sl(2) \oplus sl(2).... \oplus sl(2)}_{\times N}$ whose structure turns out to be the following

\begin{equation}
\label{sl2co}
\begin{cases}
\{ \Delta(J_3)^i , \Delta(J_+)^j \} = 2 \Delta(J_+)^i, \quad i \leq j  \\
\{ \Delta(J_3)^i , \Delta(J_-)^j \} = -2 \Delta(J_-)^i, \quad i \leq j  \\
\{ \Delta(J_-)^i , \Delta(J_+)^j \} = 4 \Delta(J_3)^i, \quad i \leq j 
\end{cases}
\end{equation}

where $i,j \leq N$. From this it is straightforward to verify that the three generators $\Delta(J_+)^N,\Delta(J_-)^N,\Delta(J_3)^N$ commute with $N$ quadratic Casimirs induced by the action of the coproduct on $\mathcal{C} = J_+ J_- - J_3^2$  

\begin{equation}
\label{ncasimir}
 \Delta(\mathcal{C})^l = \Delta(J_+)^l \Delta(J_-)^l - (\Delta(J_3)^l)^2, \quad 1 \leq l \leq N.
\end{equation}  

The elements of the above algebra can be used as fundamental bricks to express the radial Hamiltonians and its constants of the motion. Let us introduce a symplectic realization for (\ref{sl2}):

\begin{equation}
D(J_-) = x^2, \quad D(J_+) = p_x^2 , \quad D(J_3) = x p_x, \quad \{f , g \} = \sum_i \partial_i f \partial_{p_i} g - \partial_i g \partial_{p_i} f
\end{equation}

\begin{equation}
\label{rap}
D(J_-^{(N)}) = \sum_{i=1}^N x_i^2, \quad D(J_+^{(N)}) = \sum_{i=1}^N p_{x_i}^2 , \quad D(J_3^{(N)}) = \sum_{i=1}^N x_i  p_{x_i},
\end{equation}

We are finally ready to express the system (\ref{keplers}) and its constants of motions as function of the generators (\ref{rap}). Let us put

\begin{equation}
\label{traduzione}
\begin{cases}
p_\theta^2 = \mathcal{C}^{(2)} \\
r^2 = J_-^{(2)} \\
p_r = \frac{J_3^{(2)}}{\sqrt{J_-^{(2)}}} \\
p_\theta e^{- i \theta} = (xp_y - yp_x) \frac{x - iy}{r} = \frac{x \sqrt{\mathcal{C}^{(2)}}-i(x J_3^{(2)} - p_x J_-^{(2)})}{\sqrt{J_-^{(2)}}} 
\end{cases}
\end{equation}

substituting (\ref{traduzione}) into (\ref{keplers}) and (\ref{runge}) we obtain

\begin{equation}
H =  \frac{(1+k J_-^{(2)})^2}{2} (\frac{(J_3^{(2)})^2}{J_-^{(2)}} + \frac{\mathcal{C}^{(2)}}{J_-^{(2)}}) - \mu \frac{1-k J_-^{(2)}}{\sqrt{J_-^{(2)}}} +4 \mu \delta 
\end{equation}  

\begin{equation}
\label{rungeco}
S = \left( \mathcal{C}^{(2)} \frac{1-k J_-^{(2)}}{\sqrt{J_-^{(2)}}} - \mu +i (1+k J_-^{(2)}) \frac{J_3^{(2)}}{\sqrt{J_-^{(2)}}} \sqrt{\mathcal{C}^{(2)}}\right) \frac{\sqrt{\mathcal{A}}}{\sqrt{\mathcal{C}^{(2)}}}.
\end{equation}

\noindent The quantity $\sqrt{\mathcal{A}}$ is defined as:

\begin{equation}
\sqrt{\mathcal{A}} = p_\theta e^{- i \theta} = (xp_y - yp_x) \frac{x - iy}{r} = \frac{x \sqrt{\mathcal{C}^{(2)}}-i(x J_3^{(2)} - p_x J_-^{(2)})}{\sqrt{J_-^{(2)}}} .
\end{equation}

We have used the square root to emphasize that any element of the form $\alpha x + \beta p_x$ can be regarded as the square root of a function $\mathcal{F}(J_-^{(1)},J_+^{(1)},J_3^{(1)})$. 
\newpage 
\noindent For exampe we have $(\alpha x + \beta p_x)^2 = \alpha^2 J_-^{(1)} + 2 \alpha \beta J_3^{(1)} + \beta^2 J_+^{(1)}$. The complex constant of the motion (\ref{rungeco})
provides two constants of the motion given by the real and the imaginary part :

\begin{equation}
\label{realS}
Re(S) = \mathcal{C}^{(2)} \frac{(1 - k J_-^{(2)})x}{J_-^{(2)}} + \frac{(1 + k J_-^{(2)})(x J_3^{(2)}-p_x J_-^{(2)}) J_3^{(2)}}{J_-^{(2)}} - \frac{\mu x}{\sqrt{J_-^{(2)}}};
\end{equation}  

\begin{equation}
\label{imaginaryS} 
Im(S) = \frac{1}{\sqrt{\mathcal{C}^{(2)}}} \left( \frac{\mathcal{C}^{(2)} (1+k J_-^{(2)}) J_3^{(2)} x}{J_-^{(2)}} - \left( \frac{\mathcal{C}^{(2)}(1-k J_-^{(2)})}{J_-^{(2)}} - \frac{\mu}{\sqrt{J_-}} \right) (x J_3^{(2)} -p_x J_-^{(2)}) \right)
\end{equation}

\begin{equation}
\label{eq1}
\rightarrow \frac{\mathcal{C}^{(2)} (1+k J_-^{(2)}) J_3^{(2)} x}{J_-^{(2)}} - \left( \frac{\mathcal{C}^{(2)}(1-k J_-^{(2)})}{J_-^{(2)}} - \frac{\mu}{\sqrt{J_-}} \right) (x J_3^{(2)} -p_x J_-^{(2)});
\end{equation}

In (\ref{eq1})  we have multiplied the imaginary part by the constant of the motion $\sqrt{\mathcal{C}}$ to stress that the integral (\ref{rungeco}) provides two integrals of the motion polynomial in the momentum. In  particular the real part turns out to be quadratic in the momentum and in fact is the candidate to be the proper Laplace-Runge-Lenz vector.  It is possible to verify, by a direct calculation,  that $\{H(J^{(2)}),(Re(S(J^{(2)}, J^{(1)})))^2\} = 0 $ by means of the algebra (\ref{sl2co}). Since the coproduct $\Delta$ defines a homomorphism for the algebra
it is immediate to see that \newline  $\Delta^{N} \{ H(J^{(2)}),(Re(S(J^{(2)}, J^{(1)})))^2 \} =  \{ H(J^{(N)}),(Re(S(J^{(N)}, J^{(1)})))^2 \} = 0 $ holds for any dimension $N$. More explicitly we obtain that:

\begin{equation}
\label{ndimensionalham}
H = \frac{(1+k J_-^{(N)})^2}{2} (\frac{(J_3^{(N)})^2}{J_-^{(N)}} + \frac{\mathcal{C}^{(N)}}{J_-^{(N)}}) - \mu \frac{1-k J_-^{(N)}}{\sqrt{J_-^{(N)}}} +4 \mu \delta 
\end{equation}

$$
= \frac{(1 + k {\mathbf{x}}^2)^2}{2} {\mathbf{p}}^2 - \mu \frac{1 - k {\mathbf{x}}^2}{{\mathbf{x}}^2} + 4 \mu \delta .
$$

Poisson commutes with:

\begin{equation}
\label{rungeclassic}
\mathcal{L}_1 \equiv Re(S) =  \mathcal{C}^{(N)} \frac{(1 - k J_-^{(N)})x_1}{J_-^{(N)}} + \frac{(1 - k J_-^{(N)})(x_1 J_3^{(N)}-p_{x_1} J_-^{(N)}) J_3^{(N)}}{J_-^{(N)}} - \frac{\mu x_1}{\sqrt{J_-^{(N)}}}
\end{equation}

$$
= (1 - k {\mathbf{x}}^2) {\mathbf{p}}^{2} x_1 + 2 k ({\mathbf{x \cdot p}})^2 x_1 - (1 + k {\mathbf{x}}^2) ({\mathbf{x \cdot p}}) p_{x_1} - \frac{\mu x_1}{\sqrt{{\mathbf{x}}^2}}.
$$

\noindent This turns out to be the first component of the Laplace-Runge-Lenz vector of the generalized Kepler system on a N-dimensional space of constant curvature \cite{betvulpi}, which in the flat case reduces to the well known expression for the Laplace-Runge-Lenz vector on $\mathbb{E}^N$

$$
\mathcal{L}_1=  {\mathbf{p}}^{2} x_1  -  ({\mathbf{x \cdot p}}) p_{x_1} - \frac{\mu x_1}{\sqrt{{\mathbf{x}}^2}} .
$$
Finally it is possible to obtain all the other components of the Laplace-Runge Lenz vector through rotation, considering the fact that any radial Hamiltonian Poisson commutes with the set of angular momentums $L_{i,j} = x_i p_j - x_j p_i$

\begin{equation}
\mathcal{L}_i = \{a_1, L_{i,1} \} 
\end{equation}

which is consistent with the fact that in a $N$dimensional space we have $N$ possible realization for $J^{(1)}$.

\subsection{A "Laplace-Runge-Lenz" vector for higher order M.S. systems}

In the previous section we have introduced a procedure for the embedding of a two dimensional radial system in a higher dimensional space of arbitrary dimension $N$.
The crucial observation in order to obtain the constants of the motion for the entire Perlick I family
(\ref{perlickham1}) is that its projection on a two dimensional space coincides, up to the change of variables (\ref{levicivitageneral}), to the quadratic M.S. system (\ref{keplers}). This observation can be fruitfully used in the opposite direction, namely to obtain first the integrals of the motion of the 2-dimensional version of  (\ref{perlickham1}) from the integrals of (\ref{keplers}) and then to embed them in a higher dimensional space through the application of the coproduct $\Delta$. Let us show how to do this explicitly step by step.    
First let us apply the radial change of variable (\ref{levicivitageneral}) to the system (\ref{keplers}) and its two dimensionsal "Laplace-Runge-Lenz".

\begin{equation}
\label{angtransf}
\begin{cases}
\theta = \beta \theta' \\ p_\theta = \frac{1}{\beta} p_{\theta'}
\end{cases}
\end{equation}

$$
H =  \frac{(1+k r^2)^2}{2} (p_r^2 + \frac{p_\theta^2}{r^2}) - \mu \frac{1-kr^2}{r} +4 \mu \delta = \frac{(1+k r^2)^2}{2} (p_r^2 + \frac{p_{\theta}^2}{\beta^2 r^2}) - \mu \frac{1-kr^2}{r} +4 \mu \delta 
$$

$$
S =   \left( p_\theta^2 \frac{1-k r^2}{r} - \mu + i (1+kr^2) p_r p_\theta \right) e^{- i \theta} \rightarrow
$$

$$
\rightarrow \left( \frac{p_{\theta'}^2}{\beta^2} \frac{1-k r^2}{r} - \mu + i (1+kr^2) p_r \frac{p_{\theta'}}{\beta} \right) e^{- i \beta \theta'}
$$

which in terms of the coalgebra generators turn out to be

\begin{equation}
\label{perlickcoalg}
H =  \frac{(1+k J_-^{(2)})^2}{2} (\frac{(J_3^{(2)})^2}{J_-^{(2)}} + \frac{\mathcal{C}^{(2)}}{\beta^2 J_-^{(2)}}) - \mu \frac{1-k J_-^{(2)}}{\sqrt{J_-^{(2)}}} +4 \mu \delta 
\end{equation}

\begin{equation}
S = \left( \frac{\mathcal{C}^{(2)}}{\beta^2} \frac{1-k J_-^{(2)}}{\sqrt{J_-^{(2)}}} - \mu +i (1+k J_-^{(2)}) \frac{J_3^{(2)}}{\sqrt{J_-^{(2)}}} \frac{\sqrt{\mathcal{C}^{(2)}}}{\beta}\right) \left( \frac{\sqrt{\mathcal{A}}}{\sqrt{\mathcal{C}^{(2)}}} \right)^\beta .
\end{equation}

Let us remark that, in spite of the fact that the transformation (\ref{angtransf}) is just a change of variable for a two dimensional system, this induces a new coalgebraic system  (\ref{perlickcoalg}) which is indeed different from the original one (\ref{ndimensionalham}).
Moreover let us stress that after the angular transformation (\ref{angtransf}) the constant of motion $S$ is no longer polynomial in the generators $J_+,J_3$, namely the ones containing the momentum variables;  however $\beta = \frac{m}{n}$ is a rational number by hypothesis, therefore it is still possible to recover a polynomial constant of the motion considering a proper power of $S$ 

\begin{equation}
\mathcal{S} = (\sqrt{C^{(2)}})^m S^n =   \mathcal{B}^n \left( \sqrt{\mathcal{A}} \right)^m, \quad m,n \in \mathbb{N}
\end{equation}  

$$
\mathcal{B} =  \left( \frac{\mathcal{C}^{(2)}}{\beta^2} \frac{1-k J_-^{(2)}}{\sqrt{J_-^{(2)}}} - \mu +i (1+k J_-^{(2)}) \frac{J_3^{(2)}}{\sqrt{J_-^{(2)}}} \frac{\sqrt{\mathcal{C}^{(2)}}}{\beta}\right) .
$$

At this stage the constant of the motion $\mathcal{S}$ is not yet polynomial in $J_+,J_3$ since both $\mathcal{B}$ and $\sqrt{\mathcal{A}}$ depend on $\sqrt{\mathcal{C}}$, however this dependence is of the type:

$$
\mathcal{B} = (a + i b \sqrt{\mathcal{C}}), \quad \sqrt{\mathcal{A}} = (d + i f \sqrt{\mathcal{C}})
$$ 

This entails that if we consider separately the Real part and the Imaginary part of $\mathcal{S}$ we get two polynomial constants of the motion in the momentum:

\begin{equation}
\mathcal{S} = (a + i b \sqrt{\mathcal{C}})^{n} (d + i f \sqrt{\mathcal{C}})^m \rightarrow
\begin{cases}
Re(\mathcal{S}) = \mathcal{F}(J_+,J_-,J_3) \\
Im(\mathcal{S}) = \sqrt{C} \mathcal{G}(J_+,J_-,J_3) 
\end{cases}
\rightarrow \{H, \mathcal{F}\} = \{H,\mathcal{G}\} = 0 .
\end{equation}  

This provides us the way to obtain the "higher order Laplace-Runge-Lenz vector" in any N-dimensional space.
In particular let us consider the three dimensional version of (\ref{perlickcoalg})

\begin{equation}
\label{abstractham}
H =  \frac{(1+k J_-^{(3)})^2}{2} (\frac{(J_3^{(3)})^2}{J_-^{(3)}} + \frac{\mathcal{C}^{(3)}}{\beta^2 J_-^{(3)}}) - \mu \frac{1-k J_-^{(3)}}{\sqrt{J_-^{(3)}}} +4 \mu \delta 
\end{equation}

which in spherical coordinates :

\begin{equation}
\begin{cases}
x_1 = r \cos \theta \cos \phi \\
x_2 = r \cos \theta \sin \phi \\
x_3 = r \sin \theta
\end{cases} \rightarrow 
\begin{cases}
J_-^{(3)} = x_1^2 +x_2^2 +x_3^2 = r^2 \\
J_+^{(3)} = p_{x_1}^2 +  p_{x_2}^2 +  p_{x_3}^2 = p_r^2 + \frac{p_\theta^2}{r^2} + \frac{p_\phi^2}{r^2 \sin^2 \theta} \\
J_3^{(3)} = x_1 p_{x_1} + x_2 p_{x_2} + x_3 p_{x_3} = r p_r
\end{cases}
\end{equation}

$$
\mathcal{C}^{(3)} = J_+^{(3)} J_-^{(3)} -(J_3^{(3)})^2 = p_\theta^2 + \frac{p_\phi^2}{\sin^2 \theta}
$$

takes the form

\begin{equation}
H = \frac{(1+kr^2)^2}{2} \left( p_r^2 + \frac{1}{\beta^2 r^2} (p_\theta^2 + \frac{p_\phi^2}{\sin^2 \theta})\right) - \mu \frac{1-kr^2}{r} + 4 \mu \delta.
\end{equation}

This coincides with (\ref{perlickham1}) after applying the radial change of variable 

$$
r = r'^\beta
$$

\begin{equation}
H = \frac{1}{\beta^2}\left( \frac{r'^{2} (r'^{-\beta}+kr'^\beta)^2}{2} \left( p_{r'}^2 + \frac{1}{ r'^2} (p_\theta^2 + \frac{p_\phi^2}{\sin^2 \theta})\right) - \mu'(r'^{-\beta} - k r'^{\beta}) + 4 \mu' \delta \right)
\end{equation} 

where $\mu = \frac{\mu'}{\beta^2}$.

This fully agrees with the results obtained in \cite{Betal1} in which was explicitly proven the M.S. of the three dimensional systems (\ref{perlickham1}) and (\ref{perlickham2}).   

\section{Higher order M.S. systems beyond the radial symmetry}

Up to this point we have shown how the entire Perlick 1 family can be obtained from the more abstract system (\ref{abstractham}) by considering the representation (\ref{rap}). However the (\ref{rap}) is not the only possible one, in particular it is possible to get new M.S. systems by considering new representations whose symmetry is different from the radial one considered so far.
Let us choose the two dimensional representation of the algebra sl2 (\ref{rap}) 

\begin{equation}
D(J_-^{(2)}) = x_1^2 + x_2^2, \quad D(J_+^{(2)}) = p_{x_1}^2 +p_{x_2}^2, \quad D(J_3^{(2)}) = x_1 p_{x_1} + x_2 p_{x_2} .  
\end{equation}

\newpage
\noindent Let us now recast such a representation in radial variables:

\begin{equation}
\begin{cases}
x_1 = r \cos \theta, \quad p_{x_1} = \cos \theta p_r - \frac{\sin \theta}{r} p_\theta \\
x_2 = r \sin \theta, \quad  p_{x_2} = \sin \theta p_r + \frac{\cos \theta}{r} p_\theta 
\end{cases}
\end{equation}

\begin{equation}
D(J_-^{(2)}) = r^2, \quad D(J_+^{(2)}) = p_{r}^2 +\frac{p_\theta^2}{r^2}, \quad D(J_3^{(2)}) = r p_r , \quad D(\mathcal{C}^{(2)}) = p_\theta^2 .  
\end{equation}

The Casimir representation coincides with the square of the angular momentum $p_\theta$,  and the variable $\theta$ can be considered as an ignorable variable since for each algebra element holds $\partial_\theta D(J^{(2)})=0$. As a consequence we can reduce the dimension of  $D(J^{(2)})$ considering $p_\theta \rightarrow b$ as a fixed parameter  

\begin{equation}
\tilde{D}(J_-^{(1)}) = x^2, \quad \tilde{D}(J_+^{(1)}) = p_{x}^2 +\frac{b^2}{x^2}, \quad \tilde{D}(J_3^{(1)}) = x p_x  
\end{equation}

which can be considered as a generalization of the representation (\ref{rap}) when $b \neq 0$ .
Of course it is still possible to generate the two dimensional representation applying the coproduct $\Delta$ 

\begin{equation}
\label{rapttw}
\Delta (\tilde{D}(J_-^{(1)})) = x_1^2 + x_2^2 , \quad \Delta (\tilde{D}(J_+^{(1)})) = p_{x_1}^2 + p_{x_2}^2   + \frac{b_1^2}{x_1^2} + \frac{b_2^2}{x_2^2}, \quad \Delta (\tilde{D}(J_3^{(1)})) = x_1 p_{x_1} + x_2 p_{x_2}.     
\end{equation}

Let us plug in such a representation in the system (\ref{perlickcoalg}) 

$$
H =  \frac{(1+k J_-^{(2)})^2}{2} (\frac{(J_3^{(2)})^2}{J_-^{(2)}} + \frac{\mathcal{C}^{(2)}}{\beta^2 J_-^{(2)}}) - \mu \frac{1-k J_-^{(2)}}{\sqrt{J_-^{(2)}}} +4 \mu \delta \rightarrow
$$

\begin{equation}
\rightarrow \frac{(1 + k r^2)^2}{2} \left( p_r^2 + \frac{p_\theta^2}{\beta^2 r^2} + \frac{b_1^2}{\beta^2 r^2 \cos^2 \theta} + \frac{b_2^2}{\beta^2 r^2 \sin^2 \theta} \right) - \mu \frac{1 - k r^2}{r} + 4 \mu \delta .
\end{equation}

The presence of the new terms whose coupling constant is $b_i$ breaks the radial symmetry of the previous Hamiltonian obtained through the representation (\ref{rap}). If we reintroduce again the angle $\theta' = \beta \theta$ we obtain explicitly the TTW system on a space of constant curvature, which from this perspective can be regarded as the two dimensional reduction of a radial 4-dimensional system:

\begin{equation}
\label{ttwspheric}
H =  \frac{(1 + k r^2)^2}{2} \left( p_r^2 + \frac{p_{\theta'}^2}{r^2} \right) + \frac{(1 + k r^2)^2}{2} \left( \frac{b_1^2}{\beta^2 r^2 \cos^2 \beta^{-1} \theta'} + \frac{b_2^2}{\beta^2 r^2 \sin^2 \beta^{-1} \theta'} \right) - \mu \frac{1 - k r^2}{r} + 4 \mu \delta 
\end{equation}
 
which indeed is M.S. as proven in \cite{nersessian}
 
\section{The quantum version of Perlick systems}

The analysis performed on the classical radial superintegrable systems (\ref{perlickham1}), (\ref{perlickham2}) shows how the Hamiltonian and its constants of the motion can be viewed as functions of elements of the algebra (\ref{sl2co}). Its symplectic realization (\ref{rap} , \ref{rapttw})  can provide a wider family of classical superintegrable systems. The goal of this section is to take advantage of the above analysis to define a quantum realization of this algebra  with the aim of inducing, through the coalgebra approach, a superintegrable quantization for the systems (\ref{perlickham1}), (\ref{perlickham2}). 
As in the classical case, the first step in order to realize the above project, is to obtain 
a two dimensional quantum superintegrable version of (\ref{keplers}), from which, we can obtain the quantum version of (\ref{perlickcoalg}).  
\newpage
\noindent To begin with let us consider again the 2dim Kepler system (\ref{keplers}). In order to obtain its quantum analog we have to deal with the unavoidable ordering problem coming from the canonical quantization process that always arises in a non-flat space. For the Kepler system on a space of constant curvature the order ambiguity can be solved by introducing the Laplace Beltrami operator (the covariant version of the standard Laplace operator) as the quantum counterpart of the classical kinetic energy term

\begin{equation}
g^{i,j}p_i p_j \rightarrow \frac{-\hbar^2}{\sqrt{g}} \partial_i (\sqrt{g} g^{i,j} \partial_j ) .
\end{equation}

\noindent In the case of a conformally flat  two-dimensional system $ds^2 = f(r) (dr^2 + r^2 d \theta^2)$ it takes a particularly easy form which, for the sake of coherence with earlier papers, we define to be the "direct quantization", namely the quantization obtained prescribing a Hamiltonian operator selfadjoint in the Hilbert space $L^2 ( \mathbb{R}^N , f(|x|) d{\mathbf{x}}  )$ 

\begin{equation}
\label{quantdirect}
\frac{1}{f(r)} \left( p_r^2 + \frac{p_\theta^2}{r^2} \right) \rightarrow \frac{-\hbar^2}{f(r)} \nabla^2_2 = \frac{-\hbar^2}{f(r)} \left( \partial_r^2  + \frac{1}{r} \partial_r + \frac{\partial_\theta^2}{r^2}\right).
\end{equation} 

The above considerations leads to the definition of the following quantum version of (\ref{keplers}) 

\begin{equation}
\label{quantumkeplers}
\hat{H} = - \hbar^2 \frac{(1+kr^2)^2}{2} \left( \partial_r^2 + \frac{1}{r} \partial_r + \frac{\partial_\theta^2}{r^2} \right) - \mu \frac{1-kr^2}{r} .
\end{equation} 

The system (\ref{quantumkeplers}) turns out to be exactly solvable (we can calculate the spectrum and eigenfunctions) and above all it is M.S., namely it commutes with two differential operators: the angular momentum $-i \hbar\partial_\theta$ and a second differential operator which can be considered as the two-dimensional quantum version of the Laplace-Runge-Lenz vector.
In the clssical case, we obtained  the Laplace-Runge-Lenz vector from the solution of the orbit equation (\ref{orbiteq}). In the quantum case the exact solution of the bounded spectrum and its eigenfunctions can be computed algebraically factorizing the Hamiltonian operator in terms of two "ladder operators" $\hat{H} = a^\dagger a$ . On the same footing of the classical case the quantum Laplace-Runge-Lenz can be obtained by means of the same tools already used to working out its exact solution, namely using the properties of the ladder operators. The idea of using the ladder operators to get the extra integrals of the motion for a superintegrable systems is not new see i.e. \cite{kalnins} \cite{marquette2} . In the following we will adopt the same philosophy to compute the quadratical constants of the motion for the two-dimensional system with the perspective of generalizing this to the N-dimensional case as already seen in section (\ref{quattro}) for the classical case.
As a first step let us separate variables in the Schroedinger equation, i.e. putting  $\psi = e^{i l \theta} \rho (r) $

\begin{equation}
\label{radialkepler}
\hat{H}_l \rho(r) = \left(-\hbar^2 \frac{(1+kr^2)^2}{2} \left( \partial_r^2 + \frac{1}{r} \partial_r - \frac{l^2}{r^2} \right) - \mu \frac{1-kr^2}{r} \right) \rho(r) = E_l \rho(r) .
\end{equation}

The radial Schroedinger operator (\ref{radialkepler}) can be factorized introducing the operators $\hat{a}^\dagger_l , \hat{a}_l$ :

\begin{equation}
\label{ladderop}
\begin{cases}
\hat{a}^\dagger_l = \frac{1}{\sqrt{2}} \left( -i \hbar(1+ k r^2) \partial_r + i \frac{\mu}{\hbar(l+\frac{1}{2})} -i \hbar (l+1) \frac{1-kr^2}{r}  \right) \\
\hat{a}_l =  \frac{1}{\sqrt{2}} \left( -i \hbar(1+ k r^2) \partial_r -i \frac{\mu}{\hbar(l+\frac{1}{2})} + i\hbar l \frac{1-kr^2}{r}  \right) 
\end{cases}
\end{equation}

\begin{equation}
\label{factorization}
\hat{H}_l -E_l = a^\dagger_l a_l 
\end{equation}

where $E_l = -\frac{\mu^2}{2 \hbar^2 (l+ \frac{1}{2})^2} + 2k \hbar^2 (l+ \frac{1}{2})^2 - \frac{k \hbar^2}{2} . $

\newpage 
\noindent Let us stress that the operators introduced in (\ref{ladderop}) satisfy the following important relation:

\begin{equation}
\label{intertwining}
\hat{a}_l \hat{a}_l^{\dagger} + E_l = \hat{H}_{l+1} = \hat{a}^\dagger_{l+1} \hat{a}_{l+1} + E_{l+1}.
\end{equation} 

Such a property is known as shape invariance and it is an integrability condition ( for a review on SUSY QM see i.e. \cite{cooper}). It provides a way to obtain the entire discrete spectrum and the eigenfunctions.  
Let us show briefly how the shape invariance condition works.
\newline Let us consider the operator $\hat{a}^\dagger_{l+1}\hat{a}_{l+1}$, and the radial wavefunction $\rho_{0,l+1}$ which is annihilated by $\hat{a}_{l+1}$ :

\begin{equation}
\hat{a}_{l+1} \rho_{0,l+1} = 0 \rightarrow \rho_{0,l+1} = \left( \frac{r}{1+kr^2} \right)^{l+1} e^{-\frac{\mu}{\hbar^2 \sqrt{k}(l+\frac{3}{2})} \tan^{-1} (\sqrt{k}r)}
\end{equation}

the wavefunction $\rho_{0,l+1}$ can be regarded as the ground state of the Hamiltonian 
\begin{equation}
\label{eigenvalueq}
( \hat{H}_{l+1} -E_{l+1} ) \rho_{0,l+1}=\hat{a}^\dagger_{l+1} \hat{a}_{l+1} \rho_{0,l+1} = 0 .
\end{equation}

Using (\ref{intertwining}) equation (\ref{eigenvalueq}) can be recast in the form 

\begin{equation}
\hat{a}_l \hat{a}^\dagger_l \rho_{0,l+1} = (E_{l+1} - E_{l}) \rho_{0,l+1} .
\end{equation}
 
Multiplying from the left by $\hat{a}^\dagger_l$ provides the first excited state of the Hamiltonian $a^\dagger_l a_l$
 
\begin{equation}
\hat{a}^\dagger_l \hat{a}_l (\hat{a}^\dagger_l \rho_{0,l+1})  = (E_{l+1} - E_{l}) (\hat{a}^\dagger_l \rho_{0,l+1}) 
\end{equation}

or equivalently 

\begin{equation}
\hat{H}_l (\hat{a}^\dagger_l \rho_{0,l+1})  = E_{l+1} (\hat{a}^\dagger_l \rho_{0,l+1}) .
\end{equation}

It is then straightforward to obtain all the bound states and their energy eigenvalues by iteration 

\begin{equation}
\label{autofunz2d}
\rho_{n,l} \propto \prod_{j=0}^{n-1} \hat{a}^\dagger_{l+j} \rho_{0,l+n}; \quad E_{n,l} = E_{n+l}. 
\end{equation}

This provides the solution of the two dimensional eigenvalue problem (\ref{quantumkeplers})

\begin{equation}
\label{spettro}
\hat{H} \psi_{n,l} = \hat{H} e^{i l \theta} \rho_{n,l}  = \left(-\frac{\mu^2}{2 \hbar^2(l+n+ \frac{1}{2})^2} + 2k \hbar^2 (l+n+ \frac{1}{2})^2 - \frac{k \hbar^2}{2} \right) \psi_{n,l}.
\end{equation} 

Let us remark that the spectrum shows the so called "accidental degeneracy" namely it can be described by the single quantum number $\mathcal{N}=n+l$. This fact can be regarded as a direct consequence of the shape invariance property (\ref{intertwining}). The operators $\hat{a}_l$ and $\hat{a}^\dagger_l$ act as ladder operators and connect different "iso-energetic" radial eigenfunctions :

\begin{equation}
\label{ladderoperators}
\begin{cases}
\hat{a}_l \rho_{n,l} \propto \rho_{n-1,l+1} \\
\hat{a}^\dagger_l \rho_{n,l+1} \propto \rho_{n+1,l}.
\end{cases}
\end{equation}

The radial ladder operators (\ref{ladderoperators}), as introduced above,  are defined for a specific value of $l$. We can generalize their action on a generic eigenfunction $\psi_{n,l}$ by replacing the quantum number $l$ by the angular momentum $-i \hbar \partial_\theta$.
\newpage 
\noindent Let us define the operator $\hat{\tilde{S}}$:

\begin{equation}
\hat{\tilde{S}} = \frac{e^{-i \theta}}{\sqrt{2}} \left( -i \hbar (1+kr^2) \partial_r + \frac{i \mu}{-i\hbar \partial_\theta - \frac{\hbar}{2}} -\hbar \frac{1-kr^2}{r}\partial_\theta  \right).
\end{equation}

It is now just a matter of computation to see that $\hat{\tilde{S}}$ acts as a ladder operator for the set of eigenfunctions $\psi_{n,l}$ :

\begin{equation}
\hat{\tilde{S}} \psi_{n,l+1} = \frac{e^{-i \theta}}{\sqrt{2}} \left( -i \hbar (1+kr^2) \partial_r + \frac{i \mu}{-i\hbar \partial_\theta - \frac{\hbar}{2}} -\hbar \frac{1-kr^2}{r}\partial_\theta  \right) e^{i (l+1) \theta} \rho_{n,l+1} =
\end{equation}

$$
= \frac{e^{i l \theta}}{\sqrt{2}} \left( -i \hbar(1+ k r^2) \partial_r +  \frac{i \mu}{\hbar(l+\frac{1}{2})} -i \hbar (l+1) \frac{1-kr^2}{r}  \right) \rho_{n,l+1} =
$$
 
$$
=  e^{il \theta}\hat{a}^\dagger_l \rho_{n,l+1} \propto e^{il \theta} \rho_{n+1,l} = \psi_{n+1,l}
$$

it is straightforward to see that its Hermitian conjugate $\hat{\tilde{S}}^\dagger$ behaves as $\hat{a}$ :

\begin{equation}
\hat{\tilde{S}}^\dagger = \frac{e^{i \theta}}{\sqrt{2}} \left( -i \hbar (1+kr^2) \partial_r - \frac{i \mu}{-i \hbar \partial_\theta + \frac{\hbar}{2}} + \hbar \frac{1-kr^2}{r} \partial_\theta \right) 
\end{equation}

$$
\hat{\tilde{S}}^\dagger \psi_{n,l} =  \frac{e^{i \theta}}{\sqrt{2}} \left( -i \hbar (1+kr^2) \partial_r - \frac{i \mu}{-i \hbar \partial_\theta + \frac{\hbar}{2}} + \hbar \frac{1-kr^2}{r} \partial_\theta \right) e^{il \theta} \rho_{n,l} = 
$$

$$
= \frac{e^{i (l+1)\theta}}{\sqrt{2}} \left( -i \hbar (1+kr^2) \partial_r - \frac{i \mu}{\hbar( l + \frac{1}{2})} +i \hbar l \frac{1-kr^2}{r} \right) \rho_{n,l} \propto \psi_{n-1,l+1} .
$$

On the basis of the above considerations, it is natural to define the operator 

\begin{equation}
\label{quantumS}
\hat{S} = i \hat{\tilde{S}} (-i \hbar \partial_\theta - \frac{\hbar}{2}) 
\end{equation} 

which can be regarded as the quantum version of (\ref{rungeco}). Furthermore  the definition given in (\ref{quantumS}) provides a straightforward way to obtain the two dimensional quantum Laplace-Runge-Lenz vector

\begin{equation}
\hat{\mathcal{L}} = \frac{\hat{S} + \hat{S}^\dagger}{2} 
\end{equation} 

whose classical limit coincides with the (\ref{realS})

\subsection{embedding a two-dimensional radial system in a higher dimensional space}
\label{quantumembedding}
Given the quantum version of (\ref{keplers}) and its Laplace-Runge-Lenz vector (\ref{realS}) the second step consists in describing them in terms of the algebra (\ref{sl2co}). In order to accomplish to this task let us introduce the following quantum representation of  (\ref{sl2co})

\begin{equation}
[\hat{J}_3 , \hat{J}_+] = 2 i \hbar \hat{J}_+, \quad  [\hat{J}_3, \hat{J}_-] = -2 i \hbar \hat{J}_-, \quad [\hat{J}_-,\hat{J}_3] = 4i \hbar \hat{J}_3
\end{equation} 
 
 \begin{equation}
\label{sl2coquant}
\begin{cases}
[ \Delta(\hat{J}_3)^i , \Delta(\hat{J}_+)^j ] = 2i \hbar \Delta(\hat{J_+})^i, \quad i \leq j  \\
[ \Delta(\hat{J}_3)^i , \Delta(\hat{J}_-)^j ] = -2 i \hbar \Delta(\hat{J}_-)^i, \quad i \leq j  \\
[ \Delta(\hat{J}_-)^i , \Delta(\hat{J}_+)^j ] = 4i \hbar \Delta(\hat{J}_3)^i, \quad i \leq j .
\end{cases}
\end{equation}

This time the Casimir operator is  the symmetrized version of the (\ref{ncasimir})

\begin{equation}
\Delta(\hat{\mathcal{C}})^N = \frac{\Delta(\hat{J}_+)^N \Delta(\hat{J}_-^N) + \Delta(\hat{J}_-)^N \Delta(\hat{J}_+)^N}{2} - (\Delta(\hat{J}_3)^N)^2 .
\end{equation}

A quantum realization of  (\ref{sl2coquant}) is given by the following differential operators:

\begin{equation}
\label{quantumgenerators}
D(\hat{J}_-^{(N)}) = \sum_{l=1}^N x_l^2, \quad D(\hat{J}_+^{(N)}) = -\hbar^2 \sum_{l=1}^{N} \partial_{x_l}^2, \quad D(\hat{J}_3^{(N)}) = \sum_{l=1}^N (-i \hbar x_l \partial_{x_l}) - \frac{N i \hbar}{2} .
\end{equation}

Considering the quantum version of (\ref{traduzione})  

\begin{equation}
\label{traduzionequantum}
\begin{cases}
- \hbar^2 \partial_\theta^2 = \hat{p}_\theta^2 = \hat{\mathcal{C}}^{(2)} + \hbar^2 \\
r^2 = \hat{J}_-^{2} \\
- i \hbar \partial_r = \hat{p}_r = \frac{1}{\sqrt{\hat{J}_-^{(2)}}} \hat{J}_3 + \frac{i \hbar}{\sqrt{\hat{J}_-^{(2)}}} \\
e^{-i \theta} \hat{p}_\theta = \frac{x}{\sqrt{\hat{J}_-^{(2)}}} \sqrt{\hat{\mathcal{C}}^{(2)}}
 - \frac{i}{\sqrt{\hat{J}_-^{(2)}}} (x \hat{J}_3^{(2)} -\hat{J}_-^{(2)} \hat{p}_x + i \hbar x)\end{cases}
\end{equation}

it is immediate to obtain the "algebraic" version of (\ref{quantumkeplers})
   
$$
\hat{H} = - \hbar^2 \frac{(1+kr^2)^2}{2} \left( \partial_r^2 + \frac{1}{r} \partial_r + \frac{\partial_\theta^2}{r^2} \right) - \mu \frac{1-kr^2}{r}  =
$$ 

\begin{equation}
\label{quantumbetaunoperlick}
=\frac{(1+k \hat{J}_-^{(2)})^2}{2}\left(\frac{1}{\hat{J}_-^{(2)}} (\hat{J}_3^{(2)})^2 + \frac{2 i \hbar}{\hat{J}_-^{(2)}}\hat{J}_3^{(2)} + \frac{1}{\hat{J}_-^{(2)}} \hat{\mathcal{C}}^{(2)} \right) - \mu \frac{1-k \hat{J}_-^{(2)}}{\sqrt{\hat{J}_-^{(2)}}} = 
\end{equation}

$$
=\frac{(1+k \hat{J}_-^{(2)})^2}{2} \hat{J}_+^{(2)}  - \mu \frac{1-k \hat{J}_-^{(2)}}{\sqrt{\hat{J}_-^{(2)}}}
$$

and of its constant of the motion 
 
$$
\hat{\mathcal{L}} = \frac{\hat{S} + \hat{S}^\dagger}{2} \rightarrow
$$

\begin{equation}
\hat{\mathcal{L}}_i \!=\! x_i \left( \! -k \hbar^2 \!+\! (1\!- \!k \hat{J}_-^{(2)})\hat{J}_+^{(2)} \! + \! 3 i \hbar k \hat{J}_3^{(2)}  \! + \! 2k (\hat{J}_3^{(2)})^2 \! - \! \frac{\mu}{\sqrt{\hat{J}_-^{(2)}}} \! \right) \! \! - \! \left( \! (1 \! + \! k \hat{J}_-^{(2)}) \hat{J}_3^{(2)} \! + \!\frac{i \hbar}{2} \! - \! \frac{i \hbar k}{2} \hat{J}_-^{(2)} \! \right) \hat{p}_{x_i}
\end{equation}

The operator $\hat{\mathcal{L}}_i$ can be regarded as the square root of a function $\mathcal{F}(\hat{J}_+,\hat{J}_-,\hat{J}_3)$ in the sense that : 
$$\hat{\mathcal{L}}_i^2 = \mathcal{F}(\hat{J}_+,\hat{J}_-,\hat{J}_3): [\mathcal{F}(\hat{J}_+,\hat{J}_-,\hat{J}_3),\hat{H}] = 0
$$

which in view of (\ref{sl2coquant}) commutes with $\hat{H}$, as previously stressed for its classical version  (\ref{rungeclassic}). 
\newpage
\noindent It is also possible to evaluate directly the expression $[\hat{\mathcal{L}}_i, \hat{H}]$ if we enlarge the algebra (\ref{sl2coquant}) with the two elements of the Heisenberg algebra $\hat{p}_{x_i}, x_i$:

\begin{equation}
\label{heisenberg}
\begin{cases}
[x, \hat{p}_{x_i}] = i \hbar I \\
[\hat{p}_{x_i},\hat{J}_+^{(N)}] = 0, \quad \quad \quad [x_i,\hat{J}_+^{(N)}] = 2i \hbar \hat{p}_{x_i} \\
[\hat{p}_{x_i},\hat{J}_3^{(N)}] = -i \hbar \hat{p}_{x_i}, \quad \! \! \! [x_i,\hat{J}_+^{(N)}] =-i \hbar x \\
[\hat{p}_{x_i},\hat{J}_-^{(N)}] = -2i \hbar x, \quad \! [x_i,\hat{J}_-^{(N)}] = 0 .
\end{cases}
\end{equation}

This makes it straightforward to obtain the N-dimensional version of (\ref{quantumkeplers}) by choosing the $N$ dimensional realization of the algebra generators :

\begin{equation}
\hat{H}^{(N)} =\frac{(1+k \hat{J}_-^{(N)})^2}{2} \hat{J}_+^{(N)}  - \mu \frac{1-k \hat{J}_-^{(N)}}{\sqrt{\hat{J}_-^{(N)}}}
\end{equation} 

\begin{equation}
\hat{\mathcal{L}}_i^{(N)} \!\!=
\end{equation}
$$
\!\! x_i \left( \!\! -k \hbar^2 \!+\! (1\!- \!k \hat{J}_-^{(N)})\hat{J}_+^{(N)} \! + \! 3 i \hbar k \hat{J}_3^{(N)}  \! + \! 2k (\hat{J}_3^{(N)})^2 \! - \! \frac{\mu}{\sqrt{\hat{J}_-^{(N)}}} \! \right) \! \! - \! \left( \! (1 \! + \! k \hat{J}_-^{(N)}) \hat{J}_3^{(N)} \! + \!\frac{i \hbar}{2} \! - \! \frac{i \hbar k}{2} \hat{J}_-^{(N)} \! \right) \hat{p}_{x_i}
$$

\begin{equation}
[\hat{H}^{(N)},\hat{\mathcal{L}}_i^{(N)}] = 0 .
\end{equation}

\subsection{Higher order quantum "Laplace-Runge-Lenz" vector}

Carrying on the parallelism with the classical case we can get the quantization of (\ref{perlickcoalg})
from that of (\ref{keplers}) applying the substitution $\theta = \beta \theta'$  to (\ref{quantumkeplers}).  

$$
\hat{H} = - \hbar^2 \frac{(1+kr^2)^2}{2} \left( \partial_r^2 + \frac{1}{r} \partial_r + \frac{\partial_\theta^2}{r^2} \right) - \mu \frac{1-kr^2}{r} \rightarrow
$$

\begin{equation}
\label{quantumbetaperlick}
\hat{H}' = - \hbar^2 \frac{(1+kr^2)^2}{2} \left( \partial_r^2 + \frac{1}{r} \partial_r + \frac{\partial_{\theta'}^2}{ \beta ^2 r^2} \right) - \mu \frac{1-kr^2}{r} 
\end{equation}

$$
\hat{\tilde{S}} =\frac{i e^{-i  \theta}}{\sqrt{2}} \left( -i \hbar (-i \hbar\partial_\theta -\frac{\hbar}{2})(1+kr^2) \partial_r + i \mu -\hbar \frac{1-kr^2}{r} (-i \hbar\partial_\theta -\frac{\hbar}{2}) \partial_\theta  \right) \rightarrow
$$

\begin{equation}
\label{rungebetaquantum}
\hat{S}' = \frac{i e^{-i \beta \theta'}}{\sqrt{2}} \left( -i \hbar (\frac{-i \hbar}{\beta}\partial_{\theta'} -\frac{\hbar}{2})(1+kr^2) \partial_r + i \mu -\hbar \frac{1-kr^2}{\beta r} (\frac{-i \hbar}{\beta}\partial_{\theta'} -\frac{\hbar}{2}) \partial_{\theta'}  \right) .
\end{equation}

As in the classical case, this angular substitution doesn't break the M.S. whenever we consider $\beta \in \mathbb{Q}$. This fact becomes evident if we take in account how the angular substitution acts on the spectral properties of the new Hamiltonian . Let us compute the new eigenfunction:

\begin{equation}
\psi'(r,\theta')_{n,l'} = \psi(r, \beta \theta')_{n,l} = e^{i l \beta \theta'} \rho(r)_{n,l} = e^{i l' \theta'} \rho(r)_{n, \frac{l'}{\beta}}, \quad l = \frac{l'}{\beta} \rightarrow
\end{equation}

\begin{equation}
\label{spettrodegenere}
E(n+l) = E(n+ \frac{l'}{\beta}) = E(n+ \frac{m_1 l'}{ m_2}), \quad \beta= \frac{m_2}{m_1} , m_1,m_2 \in \mathbb{N}. 
\end{equation}  

\newpage
\noindent The new bound spectrum (\ref{spettrodegenere}) still exhibits the "accidental degeneracy" having an infinite set of isoenergetic eigenfunctions 

$$ < \psi_{n,l}' | \hat{H}' |\psi_{n,l}' > = < \psi_{n+s m_1 , l-s m_2}' | \hat{H}' | \psi_{n+s m_1 , l-s m_2}' >, \quad \forall s \in \mathbb{N}. $$

The operator $\hat{S}$ as defined in (\ref{rungebetaquantum}) is no longer a "ladder operator"  since it loses the property of connecting isoenergetic eigenfunctions of (\ref{quantumkeplers}), however it is possible to restore such a property redefining a new operator $_{\beta}{\hat{S}}$ as an appropriate power of $\hat{S}$  :

\begin{equation}
\label{betas}
_{\beta}{\hat{S}} \equiv \hat{S}^{m_1} = \frac{i e^{- i m_2 \theta}}{\sqrt{2}} \prod_{J=0}^{m_1-1}  \left( -i \hbar \hat{\mathcal{D}}_j (1+kr^2) \partial_r + i \mu -i  \frac{1-kr^2}{r} \hat{\mathcal{D}}_j ( \frac{-i \hbar}{\beta}\partial_\theta - \hbar j)  \right) ,
\end{equation}

$$
\hat{\mathcal{D}}_j = (\frac{-i \hbar}{\beta}\partial_\theta -\frac{\hbar}{2}(2j+1))
$$

$$
\rightarrow  {_{\beta}{\hat{S}}} \psi_{n,l} \propto \psi_{n+m_1,l-m_2}.
$$

The expression (\ref{betas}) can finally be used  to define the additional constant of the motion $_\beta\hat{\mathcal{L}}$ for (\ref{quantumbetaperlick}) 

\begin{equation}
\label{rungebetacurvo}
_\beta\hat{\mathcal{L}} = \frac{{_\beta\hat{S}}^\dagger + {_\beta\hat{S}}}{2}  .
\end{equation}

The crucial point is that the Hamiltonian operator (\ref{quantumbetaperlick}) if expressed by means of  the relations (\ref{traduzionequantum}) turns out to define a new coalgebra operator: 

\begin{equation}
\label{hamcoalgbeta}
\hat{H}^{(2)} = \frac{(1+k \hat{J}_-^{(2)})^2}{2}\left(\frac{1}{\hat{J}_-^{(2)}} (\hat{J}_3^{(2)})^2 + \frac{2 i \hbar}{\hat{J}_-^{(2)}}\hat{J}_3^{(2)} + \frac{1}{\beta^2 \hat{J}_-^{(2)}} \hat{\mathcal{C}}^{(2)}+ \frac{\hbar^2(1-\beta^2)}{ \beta^2 \hat{J}_-^{(2)}} \right) - \mu \frac{1-k \hat{J}_-^{(2)}}{\sqrt{\hat{J}_-^{(2)}}}. 
\end{equation} 

As we will prove in the following sections, the operator (\ref{hamcoalgbeta}) embedded in a $N$ dimensional space,  turns out to be exactly solvable and its bounded spectrum exhibits the so called maximal degeneracy characteristic of the M.S.quantum  systems. The proof of the M.S. for the system (\ref{hamcoalgbeta}) in any dimension $N$ requires, from a coalgebraic perspective, the evaluation of the commutator $[\hat{H}^{(2)}, {_\beta\hat{\mathcal{L}}}\left( \{\hat{J}_i^{(2)}\}, \{\hat{J}_j^{(1)}\}, \hat{p}_{x},x \right)] = 0$ in terms of the algebra  ( \ref{sl2coquant} , \ref{heisenberg}).  Such a computation is in general very hard and needs, for higher order constants of the motion ($\beta \neq 1$), several days of machine-time, however we have verified this explicitly in the case ($k \neq 0, \beta=1 ;k \neq 0 , \beta =\frac{1}{2} ;k=0, \beta = \frac{1}{2}, k=0, \beta = 2$) namely up to the third order constants of the motion.  
If we conjecture that the M.S. holds for any value of the rational parameter $\beta$ than it is straightforward to obtain the explicit expression of the constants of the motion in any dimension $N$. Let us prove this considering for simplicity the square of the Laplace-Runge-Lenz ${_\beta\hat{\mathcal{L}}}^2 = \mathcal{G} \left( \{\hat{J}_i^{(2)}\}, \{\hat{J}_j^{(1)}\} \right)$, which in analogy with the classical case can be expressed by using only the generators of (\ref{sl2coquant}).

\begin{equation}
\label{commutazionebeta}
[H(\{\hat{J}_i^{(2)}\}), \mathcal{G}(\{\hat{J}_i^{(2)}\}, \{\hat{J}_j^{(1)}\}) ] = 0 \rightarrow [H(\{\hat{J}^{(N)}\}), \mathcal{G}(\{\hat{J}^{(N)}\}, \{\hat{J}^{(m)}\}) ] = 0, \quad m \leq N.
\end{equation}

\newpage
\noindent Let us remark that if we set $m=1$ then there are $N$ possible realization for the set of elements $\{\hat{J}_i^{(1)}\}$ namely 

$$
\hat{J}_+^{(1)} = -\hbar^2 \partial_{x_l}^2, \quad \hat{J}_+^{(1)} = -i \hbar x_l \partial_{x_l} - \frac{i \hbar}{2}, \quad \hat{J}_-^{(1)} = x_l^2, \quad l=1,2,....N
$$

providing N constants of the motion ($\mathcal{G}(\{\hat{J}^{(N)}\}, \{\hat{J}^{(1)}\}_l)$) which, together with the set of angular momenta, makes the quantum system (\ref{quantumbetaperlick}) a M.S. Hamiltonian.    

\subsection{spectrum and eigenfunctions of the n-dimensional system (\ref{quantumbetaperlick}) }

Let us conclude the analysis on the N-dimensional embedding of (\ref{quantumbetaperlick}) evaluating explicitly its eigenfunctions and the spectrum 

\begin{equation}
\label{hn}
\hat{H}^{(N)} = \frac{(1+k \hat{J}_-^{(N)})^2}{2}\left(\frac{1}{\hat{J}_-^{(N)}} (\hat{J}_3^{(N)})^2 + \frac{2 i \hbar}{\hat{J}_-^{(N)}}\hat{J}_3^{(N)} + \frac{1}{\beta^2 \hat{J}_-^{(N)}} \hat{\mathcal{C}}^{(N)}+ \frac{\hbar^2(1-\beta^2)}{ \beta^2 \hat{J}_-^{(N)}} \right) - \mu \frac{1-k \hat{J}_-^{(N)}}{\sqrt{\hat{J}_-^{(N)}}} 
\end{equation} 

let us introduce the N-dimensional radial coordinates :

\begin{equation}
\label{radialvariables}
x_j = r \cos \theta_j \prod_{k=1}^{j-1} \sin \theta_k ; \quad x_N = r \prod_{k=1}^{N-1} \sin \theta_k
\end{equation} 

consequently the generators (\ref{quantumgenerators}) become 

\begin{equation}
\label{radialrapn}
\hat{J}_-^{(N)} = r^2, \quad \hat{J}_+^{(N)}=  -\hbar^2 \partial_r^2 - \frac{\hbar^2(N-1)}{r}\partial_r + \frac{\hat{L}^2}{r^2}, \quad \hat{J}_3^{(N)} = -i \hbar r \partial_r - \frac{i \hbar N}{2}
\end{equation}

$$
\hat{L}^2 = \sum_{j=1}^{N-1} \left( \prod_{k=1}^{j-1} \frac{1}{\sin^2 \theta_k}\right) \frac{1}{(\sin \theta_j)^{N-1-j}} (-i \hbar \partial_{\theta_j}) (\sin \theta_j)^{N-1-j} (-i \hbar \partial_{\theta_j}).
$$

Using the radial representation (\ref{radialrapn}) the Hamiltonian operator (\ref{hn}) takes the form 

\begin{equation}
\hat{H}^{(N)} =  \frac{(1+k r^2)^2}{2}\left(-\hbar^2 \partial_r^2 -\hbar^2 \frac{N-1}{r}\partial_r + \frac{\hat{L}^2}{\beta^2 r^2} + \frac{\hbar^2}{4 r^2} \left( \frac{1}{\beta^2} -1 \right) (N-2)^2   \right) -\mu \frac{1-k r^2}{r}.
\end{equation}

Let us separate variables in the eigenfunction  $\psi(r, {\mathbf{\theta}})_{n,l,l_1...l_{N-2}}= \rho(r)_{n,l} Y({\mathbf{\theta}})_{l,l_1...l_{N-2}}$. The functions $Y({\mathbf{\theta}})_{l,l_1...l_{N-2}}$ are the set of hyperspherical harmonics functions, namely the eigenfunctions of $\hat{L}^2$ which satisfy the eigenvalue equation given by:

$$
\hat{L}^2 Y({\mathbf{\theta}})_{l,l_1...l_{n-2}} = \hbar^2 l(l+N-2) Y({\mathbf{\theta}})_{l,l_1...l_{N-2}}.
$$

The above factorization leads to the radial Hamiltonian operator:

\begin{equation}
\label{hn}
\hat{H}_r^{(N)} =  \frac{(1+k r^2)^2}{2}\left(-\hbar^2 \partial_r^2 -\hbar^2 \frac{N-1}{r}\partial_r + \frac{\hbar^2 l(l+N-2)}{\beta^2 r^2} + \frac{\hbar^2}{4 r^2} \left( \frac{1}{\beta^2} -1 \right) (N-2)^2   \right) -\mu \frac{1-k r^2}{r}
\end{equation}

however the operator $\hat{H}_r^{(N)}$ is gauge equivalent to (\ref{quantumkeplers}) 

\begin{equation}
r^{\frac{N-2}{2}} \hat{H}_r^{(N)} r^{\frac{2-N}{2}} =   \frac{(1+k r^2)^2}{2}\left(-\hbar^2 \partial_r^2 -\hbar^2 \frac{1}{r}\partial_r + \frac{\hbar^2 \tilde{l}^2}{r^2}   \right) -\mu \frac{1-k r^2}{r}, \quad \tilde{l} = \frac{l + \frac{N-2}{2}}{\beta}
\end{equation} 

considering the expressions (\ref{autofunz2d}) and (\ref{spettro}) we can finally determine the spectrum of (\ref{hn})

\begin{equation}
E_{n,l} =  \left(-\frac{\mu^2}{2 \hbar^2(\frac{l}{\beta}+n + \frac{N-2}{2 \beta} + \frac{1}{2})^2} + 2k \hbar^2 (\frac{l}{\beta}+n + \frac{N-2}{2 \beta} + \frac{1}{2})^2 - \frac{k \hbar^2}{2} \right).
\end{equation} 

Before concluding the section let us also remark that the Hamiltonian (\ref{hn}) is indeed equivalent to the direct quantization of the N-dimensional Perlick I system defined in (\ref{quantdirect}). 
If we define the new radial variable $r= r'^{\beta}$ and apply the gauge transformation $r'^{\frac{(2-N)(1-\beta)}{2}} \hat{H}^{(N)} r'^{\frac{(N-2)(1-\beta)}{2}}$ we obtain the new Hamiltonian operator:

\begin{equation}
\label{ndimdirectquant}
r'^{\frac{(2-N)(1-\beta)}{2}} \hat{H}^{(N)} r'^{\frac{(N-2)(1-\beta)}{2}} = - \hbar^2 \frac{r'^2 (r'^{-\beta} +k r'^{\beta})^2}{2 \beta^2} \nabla^2_N -\mu (r'^{-\beta} -k r'^{\beta}). 
\end{equation}

This means that the coalgebra symmetry induces the "direct quantization" and for this reason the direct quantization is the quantization which keeps the M.S. propety for the Perlick systems as already stressed in \cite{Betal7}, \cite{Rig}.  On the other hand if we demand a covariant quantization we have to consider a quantum correction term to the potential proportional to the scalar curvature of the system which makes the Laplace Beltrami quantization of the N-dimensional system (\ref{perlickham1}) gauge equivalent to (\ref{ndimdirectquant}) and therefore M.S. .

\begin{equation}
\hat{H}_{L.B.} = -\frac{\hbar^2}{2} \frac{1}{\sqrt{g}} \partial_i \left( \sqrt{g} g^{i,j} \partial_j \right) + V(r) + \hbar^2 \frac{N-2}{8(N-1)} R(r)
\end{equation}

\begin{equation}
f^{\frac{2-N}{4}} \hat{H}_{d} f^{\frac{N-2}{4}} = \hat{H}_{L.B.}
\end{equation}

where $f = \frac{1}{r^2 (r^{-\beta} +k r^{\beta})^2}$

\section{4-dimensional Perlick quantum systems and TTW systems }
We conclude the paper by obtaining explicitly the generalization of the quantum TTW system on a space of constant scalar curvature by reduction of the 4-dimensional Perlick quantum system:

\begin{equation}
\label{hn4}
\hat{H}^{(4)} = \frac{(1+k \hat{J}_-^{(4)})^2}{2}\left(\frac{1}{\hat{J}_-^{(4)}} (\hat{J}_3^{(4)})^2 + \frac{2 i \hbar}{\hat{J}_-^{(4)}}\hat{J}_3^{(4)} + \frac{1}{\beta^2 \hat{J}_-^{(4)}} \hat{\mathcal{C}}^{(4)}+ \frac{\hbar^2(1-\beta^2)}{ \beta^2 \hat{J}_-^{(4)}} \right) - \mu \frac{1-k \hat{J}_-^{(4)}}{\sqrt{\hat{J}_-^{(4)}}} .
\end{equation} 

Following the same strategy shown in section 5 let us compute the reduction of the 4-dimensional realization of (\ref{quantumgenerators}) .

\begin{align}
\hat{J}_-^{(4)} & = x_1^2 + x_2^2 + x_3^2 + x_4^2 ; \\
\hat{J}_3^{(4)} &= -i \hbar (x_1 \partial_{x_1} + x_2 \partial_{x_2} + x_3 \partial_{x_3} + x_4 \partial_{x_4} + 2) ; \\
\hat{J}_+^{(4)} & = -\hbar^2 (\partial_{x_1}^2 +\partial_{x_2}^2 + \partial_{x_3}^2 + \partial_{x_4}^2) .
\end{align}

\newpage 
\noindent This representation can be reduced to a lower dimensional one by using a bi-polar coordinates system:

\begin{align}
\label{variabiliridotte}
x_1 = r_1 \cos \phi_1 & \quad ; \quad x_2 = r_1 \sin \phi_1 \\ \nonumber
x_3 = r_2 \cos \phi_2 & \quad ; \quad x_4 = r_2 \sin \phi_2  
\end{align}

\begin{align}
\hat{J}_-^{(4)} & = r_1^2 + r_2^2 ; \\
\hat{J}_3^{(4)} &= -i \hbar (r_1 \partial_{r_1} + r_2 \partial_{r_2} + 2) ; \\
\hat{J}_+^{(4)} & = -\hbar^2 (\partial_{r_1}^2 + \frac{1}{r_1} \partial_{r_1} + \frac{1}{r_1^2} \partial_{\phi_1}^2  + \partial_{r_2}^2 + \frac{1}{r_2} \partial_{r_2} +\frac{1}{r_2^2} \partial_{\phi_2}^2 ).
\end{align}

Since the generators are independent on the new angular variables $\phi_1$ , $\phi_2$ it is possible 
to get rid of the two degrees of freedom coming from $\phi_1,\phi_2$ and to obtain a new two dimensional system which inherits the properties of the original 4-dimensional one.
Let us show in a few algebraical steps how to perform the reduction for the quantum case

\begin{equation}
\label{version4d}
< \psi(x_1,x_2,x_3,x_4) | \hat{H}_{4d} | \psi (x_1,x_2,x_3,x_4) > =
\end{equation}

$$
= \int f \psi(x_1,x_2,x_3,x_4)^* \hat{H}_{4d} \psi (x_1,x_2,x_3,x_4) dx_1 dx_2 dx_3 dx_4 =
$$

$$
= \int f r_1 r_2 \psi(r_1,r_2, \phi_1, \phi_2)^* \hat{H}_{4d} \psi(r_1,r_2, \phi_1, \phi_2) dr_1 dr_2 d \phi_1 d \phi_2 
$$

the next step is to separate the wave function $ \sqrt{r_1 r_2} \psi (r_1,r_2,\phi_1,\phi_2) = \tilde{\psi} (r_1, r_2) e^{i l_1 \phi_1} e^{i l_2 \phi_2} $ so that the (\ref{version4d}) turns into:

\begin{equation}
\int f \tilde{\psi} (r_1, r_2)^* \hat{\tilde{H}}_{2d} \tilde{\psi} (r_1, r_2) dr_1 dr_2 
\end{equation}

where the reduced operator is defined by:

\begin{equation}
\hat{\tilde{H}}_{2d} = \sqrt{r_1 r_2}\left( \frac{1}{4 \pi^2} \int e^{-i l_1 \phi_1} e^{-i l_2 \phi_2} \hat{H}_{4d} e^{i l_1 \phi_1} e^{i l_2 \phi_2}  d\phi_1 d \phi_2 \right) \frac{1}{\sqrt{r_1 r_2}}.
\end{equation}

The reduced Hamiltonian can be now written in terms of the reduced version of the original generators $$\hat{J}_i^{(4)}\rightarrow \hat{\tilde{J}}_i^{(4)} = \sqrt{r_1 r_2}(\frac{1}{4 \pi^2} \int e^{-i l_1 \phi_1} e^{-i l_2 \phi_2} \hat{J}_i^{(4)} e^{i l_1 \phi_1} e^{i l_2 \phi_2} d \phi_1 d \phi_2 ) \frac{1}{\sqrt{r_1 r_2}}$$ 

\begin{align}
\label{nonradialqr}
\hat{\tilde{J}}_-^{(4)} & = r_1^2 + r_2^2 \equiv \hat{J}_-^{(2)} ; \\
\hat{\tilde{J}}_3^{(4)} &= -i \hbar (r_1 \partial_{r_1} + r_2 \partial_{r_2} + 1) \equiv \hat{J}_3^{(2)}; \\
\hat{\tilde{J}}_+^{(4)} & = -\hbar^2 (\partial_{r_1}^2 + \frac{1-4 l_1^2}{4 r_1^2}   + \partial_{r_2}^2  + \frac{1-4l_2^2}{4 r_2^2} ) \equiv \hat{J}_+^{(2)} + \hbar^2 \frac{4 l_1^2 - 1}{4 r_1^2} + \hbar^2 \frac{4l_2^2 -1 }{4 r_2^2}.
\end{align}

After the reduction, the 4-dimensional representation coincides with the 2-dimensional ones except for the generator $\hat{J}_+$ wich has an additional centrifugal term depending on the quantum numbers $l_1, l_2$ coming from the degrees of freedom we had cut off previously.
Eq. (\ref{nonradialqr}) can be regarded as a quantum non radial representation of the algebra (\ref{sl2coquant}). If we plug this representation into the "algebraic" Perlick Hamiltonian (\ref{hn}) we obtain the system :

\begin{equation}
\hat{H}^{(2)} = \frac{(1+kr^2)^2}{2} \left( -\hbar^2 \partial_{r}^2 -\hbar^2 \frac{1}{r}\partial_r -\hbar^2 \frac{1}{r^2 } \partial_{\theta'}^2+ \frac{b_1}{r^2 \cos^2 \frac{\theta'}{\beta}} + \frac{b_2}{r^2 \sin^2 \frac{\theta'}{\beta}} \right) - \mu \frac{1-kr^2}{r}
\end{equation}

where we have expressed the generators (\ref{nonradialqr}) in terms of the variables

$$
r_1 = r \cos \frac{\theta'}{\beta}
$$

$$
r_2 = r \sin \frac{\theta'}{\beta}.
$$

We have reabsorbed the "old" quantum numbers "$l_1$" , "$l_2$" in the parameters  $b_1 = \frac{1-4l_1^2}{4 \beta^2}$, $b_2 = \frac{1-4l_2^2}{4 \beta^2}$. This is a generalization of the the TTW system to spaces of constant scalar curvature.

\section{Conclusions}

The main results of this paper can be summed up in the following three points :

\noindent 1) We have introduced a technique based on the coalgebra to generate M.S. classical and quantum systems in N-dimensions from 2-dimenionl ones. 
In particular we have shown how the family of 2-dimensional radial M.S. systems with quadratic constants of the motion can generate superintegrable systems with higher order constants of the motion of arbitrary order by immersion in spaces of dimension $N>2$.     
As a particular case we have obtained the 3-dimensional Perlick classification of classical Bertrand-Hamiltonians by embedding of the two dimensional systems Darboux III and Darboux IV.
\newline 2) All the systems analyzed in the present paper have been considered both in their classical and quantum mechanical version. In particular we stress the connection with the SUSY Quantum Mechanics. The radial Schroedinger equation associated to this class of superintegrable systems can be exactly solved by mean of the shape invariance technique and moreover all the integrals of the motion can be straightforwardly obtained through the embedding of the ladder operators in a higher dimensional space.  
\newline 3)  The connection with the family of TTW systems is also discussed. We show that the family of radial M.S. systems and the TTW share the same coalgebra and indeed the TTW can be regarded as a dimensional reduction of a 4-dimensional M.S. radial system inheriting in this way all its integrability and solvability properties.

 \section{Acknowledgments}
The author acknowledges a fellowship from the laboratory of Mathematical Physics of CRM.    The author wishes also to thank Professor P. Winternitz for his important comments and insights.   

\end{document}